\documentclass[9pt,twocolumn,twoside]{opticajnl}
\journal{opticajournal} 

\setboolean{shortarticle}{true}

\usepackage{multicol}
\usepackage{lineno}
\usepackage{amsmath}
\usepackage{amssymb}
\newcommand{\abs}[1]{\ensuremath{\left| #1 \right|}}

\title{Hermite-Laguerre-Gaussian Vector Modes}

\author[1]{Edgar Medina-Segura}
\author[1]{Leonardo Miranda-Culin}
\author[2]{Benjamin Perez-Garcia}
\author[1*]{Carmelo Rosales-Guzm\'an}
\author[3*]{Mitchell A. Cox}
\affil[1]{Centro de Investigaciones en Óptica, A.C., Loma del Bosque 115, Colonia Lomas del campestre, 37150 León, Gto., Mexico.}
\affil[2]{Photonics and Mathematical Optics Group, Tecnologico de Monterrey, Monterrey 64849, Mexico}
\affil[3]{School of Electrical and Information Engineering, University of the Witwatersrand, Johannesburg, South Africa}
\affil[*]{Corresponding author's: mitchell.cox@wits.ac.za; carmelorosalesg@cio.mx}

\begin{abstract} 
Vector modes are well-defined field distributions with spatially varying polarisation states, rendering them irreducible to the product of a single spatial mode and a single polarisation state. Traditionally, the spatial degree of freedom of vector modes is constructed using two orthogonal modes from the same family. In this letter, we introduce a novel class of vector modes whose spatial degree of freedom is encoded by combining modes from both the Hermite- and Laguerre-Gaussian families. This particular superposition is not arbitrary, and we provide a detailed explanation of the methodology employed to achieve it. Notably, this new class of vector modes, which we term Hybrid Hermite-Laguerre-Gaussian (HHLG) vector modes, gives rise to subsets of modes exhibiting polarisation dependence on propagation due to the difference in mode orders between the constituent Hermite- and Laguerre-Gaussian modes. To the best of our knowledge, this is the first demonstration of vector modes composed of two scalar modes originating from different families. We anticipate diverse applications for HHLG vector modes in fields such as free-space communications, information encryption, optical metrology, and beyond.
\end{abstract}

\begin{document}
\maketitle

The Helmholtz paraxial wave equation, when solved in various coordinate systems, unveils shape-invariant families of optical fields that propagate freely in space. Specifically, the common scalar solutions with uniform polarisation—Hermite-Gaussian (HG), Laguerre-Gaussian (LG), and Ince-Gaussian (IG) modes—are derived in Cartesian, polar, and elliptical coordinates, respectively \cite{saleh2019fundamentals,IG_2004}. These modes collectively form complete and orthonormal bases within an infinite Hilbert space \cite{ForbesBook2014}, enabling any paraxial optical field to be represented as a complex superposition of these scalar modes.

The polarisation of light, which occupies a two-dimensional vector space, typically involves linear and circular polarisations as its primary bases. Unlike traditional modes, pure vector modes of light exhibit varying polarisation states across their transverse plane and thus cannot be simply reduced to the product of a spatial mode and a polarisation vector. Such vector modes are crafted through the superposition of two orthogonal optical fields (scalar modes) with corresponding orthogonal polarisation states \cite{Rosales-Guzman_2018}. Following this principle, Hermite- and Laguerre-Gaussian vector modes \cite{Maurer_2007}, as well as other innovative structures like vector Bessel beams \cite{vectorBessel_2013}, Airy vector beams \cite{airyVector_2015}, "classically entangled" Ince-Gaussian modes \cite{yaoli2020}, Helical Mathieu-Gauss vector modes \cite{Mathieu_VB_2021}, Parabolic vector beams \cite{parabolicVB_2021}, Parabolic accelerating vector waves \cite{parabolicAccVB_2021}, and the recent Helico-Conical Vector Beams \cite{HCVB_Medina-Segura_2023}, have been developed. These vector beams showcase distinctive properties that are leveraged in high-speed kinematic sensing \cite{appKinematic_2015}, holographic optical trapping \cite{optTrap_2018}, visible light communications \cite{optComm_2015}, mode division multiplexing \cite{cox2020slt}, resilience against turbulence \cite{Cox2016,Peters2023}, and quantum optics communications \cite{ndagano2017b,ndagano2017characterizing}.

\begin{figure}[tb]
    \centering
    \includegraphics[width=0.49\textwidth]{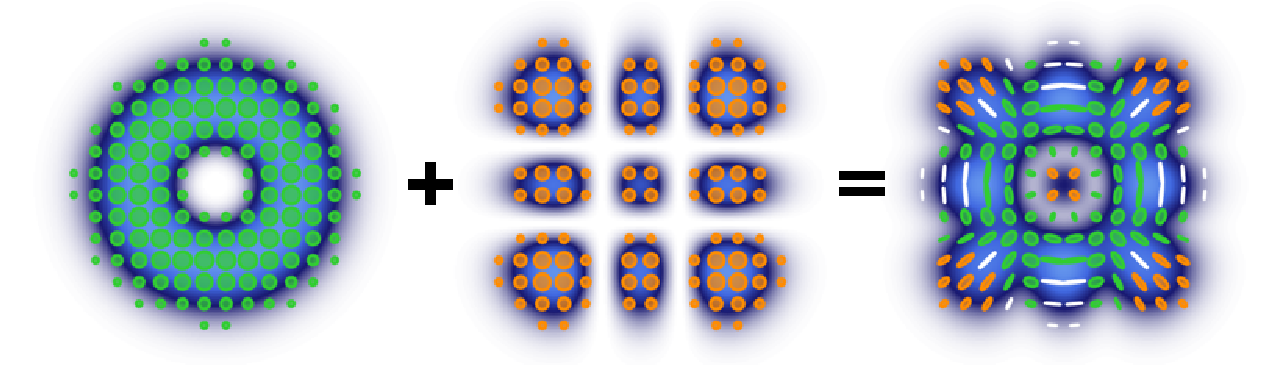}
    \caption{Graphic representation of a HHLG vector mode employing $LG^2_0$ and $HG_{2,2}$ modes, with $\theta=\pi/4$ and $\alpha=0$. Green ellipses illustrate right circular polarisation, orange ellipses depict left circular polarisation, and white lines indicate linear polarisation states.}
    \label{img_VBconcept}
\end{figure}

Typically, the scalar optical fields that constitute these vector modes originate from the same family. However, this conventional approach is not necessarily a strict requirement. In our work, we introduce a novel class of optical vector modes, which we have named Hybrid Hermite-Laguerre-Gaussian (HHLG) vector modes, by combining two scalar modes from distinct families. Specifically, we use HG and LG modes, though our method can be applied to other modal bases with suitable properties. This superposition, incorporating orthogonal polarisation states from the circular polarisation basis, exploits a well-established expression that links HG and LG modes through their completeness property to identify compatible pairs of scalar modes \cite{OC_Beijers_1993,OC_ONeil_2000,PRA_Cox_2018}. This technique enables the creation of novel spatial phase, intensity, and polarisation structures with stable propagation characteristics. Additionally, these structures can be engineered to exhibit polarisation dependence on propagation, due to differences in mode orders between the constituent modes, with an example shown in Fig.~\ref{img_VBconcept}.

\medskip

\noindent The general mathematical expression for a vector beam $\Vec{U}(\Vec{r}_\perp,z)$ is as follows:
\begin{equation}
    \Vec{U}(\Vec{r}_\perp,z)=\cos\theta\:\:u_1(\Vec{r}_\perp,z)\hat{e}_R+e^{i\alpha}\sin\theta\:\:u_2(\Vec{r}_\perp,z)\hat{e}_L,
    \label{ec_vectorBeam}
\end{equation}
where $u_1(\Vec{r}_\perp,z)$ and $u_2(\Vec{r}_\perp,z)$ are two orthogonal scalar optical fields, $\hat{e}_R$ and $\hat{e}_L$ are the circular right- and left-handed unitary polarisation vectors, $\theta$ is a weighting factor, $\alpha$ is the inter-modal phase, $z$ is the propagation axis and $\Vec{r}_\perp=(x,y)=(\rho,\phi)$ represents the transverse coordinate system for $u_1$ and $u_2$. For the vector modes that we propose, $u_1(\Vec{r}_\perp,z)$ is an optical field from the Laguerre-Gaussian basis and $u_2(\Vec{r}_\perp,z)$ is from the Hermite-Gaussian basis, or vice versa. Thus, the mathematical expression for the HHLG vector modes is given by 
\begin{equation}
    \Vec{H}_{n,m}^{\ell,p}(\Vec{r}_\perp,z)=\cos\theta\:\:LG^{\ell}_p(\rho, \phi,z)\hat{e}_R+e^{i\alpha}\sin\theta\:\:HG_{n,m}(x,y,z)\hat{e}_L,
    \label{ec_vectorHybridBeam}
\end{equation}
where the $HG_{n,m}$ and $LG^{\ell}_p$ are orthogonal Hermite- and Laguerre-Gaussian modes. We will describe how to find these combinations later. In Cartesian coordinates, $(x,y,z)$, the HG modes are defined as
\begin{equation}
    \begin{split} 
    HG_{n,m}(x,y,z) = & \sqrt{\frac{2}{\pi}}2^{-(m+n)/2} \frac{1}{\sqrt{n!m!w_0^2}} \times \\ 
                     & H_n\left[\sqrt{2}x/w(z)\right] H_m\left[\sqrt{2}y/w(z)\right] \times \\
                     & u_G(x,y,z) \exp[i\Psi_H(z)],
\end{split}
\end{equation}
where $m$, $n$ are the indices of the mode, $H_m(\cdot)$ is the Hermite polynomial of order $m$, $w(z) = w_0\sqrt{1 + (z/z_r)^2}$, $z_r = \pi w_0^2/\lambda$ and $w_0$ is the beam waist at the plane $z=0$. $\Psi_H(z)$ is a propagation-dependent phase shift (described later), and 
\begin{equation}
    \begin{split} 
    u_G(x,y,z) = & \frac{1}{\sqrt{1+(z/z_r)^2}} \exp\left[ik(x^2 + y^2)/2R(z)\right] \times \\
                 & \exp\left[-(x^2 + y^2)/w^2(z)\right]
\end{split}
\end{equation}
represents a Gaussian term, with $k=2\pi / \lambda$ and $R(z) = z+z_r^2/z$.

Similarly, in polar cylindrical coordinates, $(\rho, \phi, z)$, the LG mode can be expressed as
\begin{equation}
    \begin{split}
    LG_{p}^{\ell}(\rho,\phi,z) = & \sqrt{\frac{2n!}{\pi w_0^2 (n +\abs{\ell})!}} \left(\frac{\sqrt{2}\rho}{w(z)}\right)^{\abs{\ell}} \times \\
                            & \mathcal{L}_p^{\abs{\ell}}\left(\frac{2\rho^2}{w^2(z)}\right) u_G(\rho, \phi, z) \times \\
                            & \exp(i\ell\phi) \exp[i\Psi_L(z)],
\end{split}
\end{equation}
where $\mathcal{L}_p^{\abs{\ell}}(\cdot)$ is the Laguerre polynomial, $p$ is the radial index and $\ell$ is known as the topological charge.  The phase term $\Psi_L(z)$ is a propagation dependent phase shift.


To create the HHLG vector modes we must select modes from each family that are orthogonal. In order to identify orthogonal pairs of modes from the different families we begin our analysis by expressing Laguerre-Gaussian ($LG^{\ell}_p$) modes as linear combinations of Hermite-Gaussian ($HG_{n,m}$) modes \cite{OC_Beijers_1993,OC_ONeil_2000,PRA_Cox_2018},
\begin{equation}
    LG_{p}^l(\rho,\phi,z) = \sum_{k=0}^{N}i^{k}b(n,m,k)\mbox{HG}_{N-k,k}(x,y,z),
    \label{ec_LGtoHG}
\end{equation}
obeying the index relations $n=(2p+|\ell|+\ell)/2$ and $m=(2p+|\ell|-\ell)/2$. The coefficients $b(n,m,k)$ of the superposition are given by
\begin{equation}
    b(n,m,k)=\sqrt{\frac{(N_*-k)!k!}{2^{N_*}n!m!}}\frac{1}{k!}\left[\frac{d^k(1-t)^n(1+t)^m}{dt^k}\right]_{t=0},
    \label{ec_b}
\end{equation}
where $N_L=2p+|\ell|$ and $N_H=n+m$ is the mode order (generically $N_*$). The beam propagation factor, which describes the divergence of a beam, is simply $M^2=N_*+1$.

For different mode order, orthogonality is guaranteed; however, there are cases where orthogonality occurs within the same $N_*$. For a given $LG^{\ell}_p$ mode, we first identify the equivalent combination of $HG_{n,m}$ modes and then select the remaining modes with the same order, i.e. where $b=0$.

By way of example, we express the $LG^{-2}_1$ as a linear combination of $HG_{n,m}$ modes by using (\ref{ec_LGtoHG}) and (\ref{ec_b}) as:
\begin{equation}
    LG^{-2}_1=\frac{1}{2}HG_{4,0}+\frac{i}{2}HG_{3,1}+\frac{i}{2}HG_{1,3}-\frac{1}{2}HG_{0,4}.
    \label{ec_linearCombinationLG}
\end{equation}
Thus, we can conclude that $LG^{-2}_1$ is \textit{not} orthogonal to $HG_{4,0}$, $HG_{3,1}$, $HG_{1,3}$, or $HG_{0,4}$ because it is a linear combination of these modes. However, the remaining $N_*=4$ mode, $HG_{2,2}$, must be orthogonal to $LG^{-2}_1$ as it does not appear in (\ref{ec_linearCombinationLG}). 

\begin{figure}[tb]
    \centering
    \includegraphics[width=0.4\textwidth]{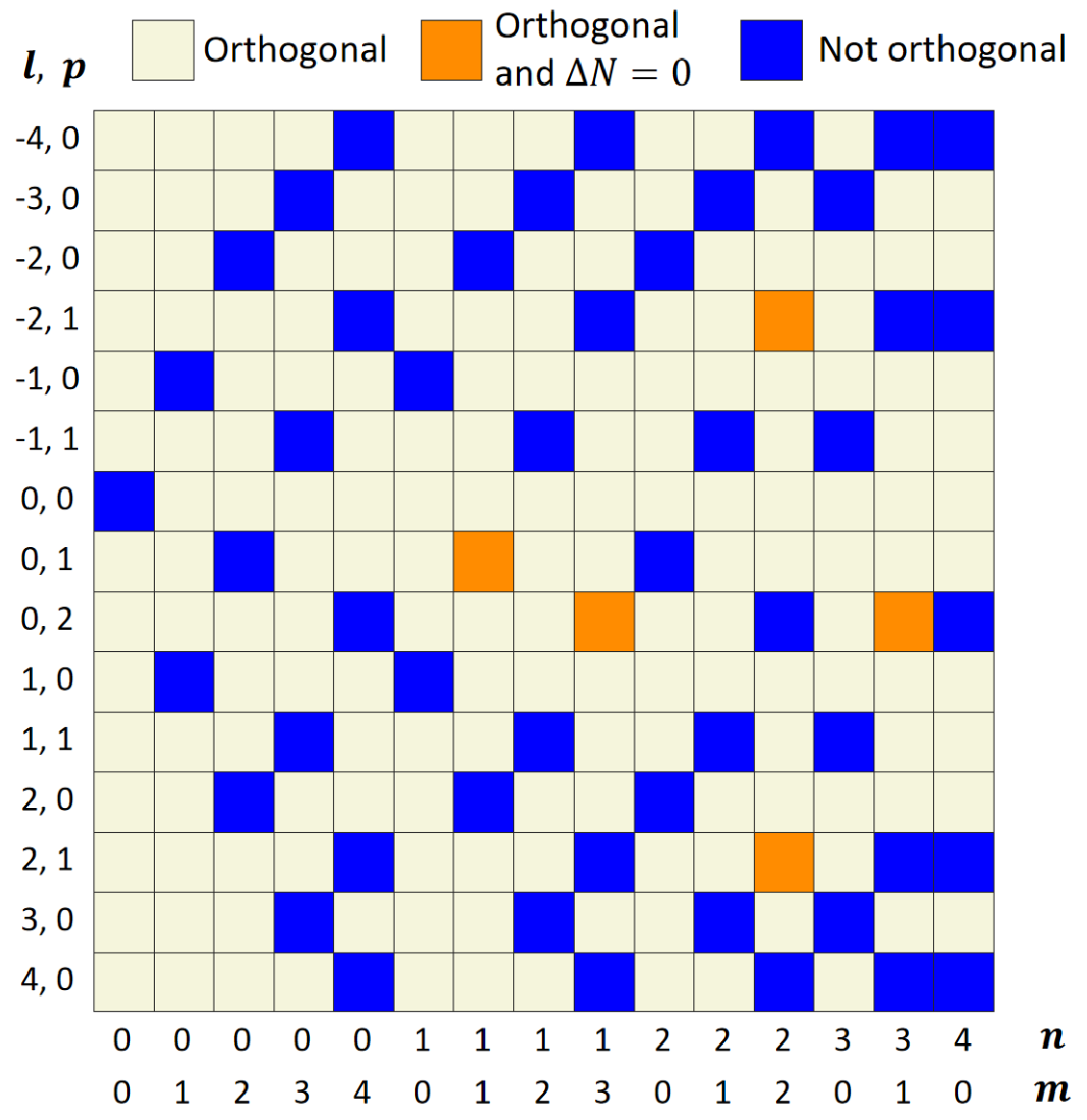}
    \caption{For $LG^{l}_p$ and $HG_{nm}$ modes with $N_L, N_H \leq 4$ there are 225 possible combinations. From these, 175 pairs of modes show orthogonality (beige) and five of them have the same mode order (orange).}
    \label{img_matrixOrthogonal}
\end{figure}

In other words, any $LG^{\ell}_p$ is orthogonal to any $HG_{n,m}$ that does not appear in its Hermite-Gauss linear decomposition. Without loss of generality, we have restricted our analysis to a subset of modes where $N_* \leq 4$, encompassing fifteen $HG_{n,m}$ modes and fifteen $LG^{\ell}_p$ modes. This leads to 225 possible combinations of modes as shown in Fig.~\ref{img_matrixOrthogonal}. Within these combinations, 175 satisfy the orthogonality condition (beige colour), including five pairs that share the same mode order (orange colour). These five combinations, crucial to this new family of vector modes, are expected to preserve their non-homogeneous polarisation patterns during propagation. Conversely, the polarisation structure of the remaining 170 vector modes is anticipated to evolve predictably with propagation. The rate of this evolution will depend on the mode order difference, as is detailed below.

\begin{figure}[tb]
    \centering
    \includegraphics[width=0.4\textwidth]{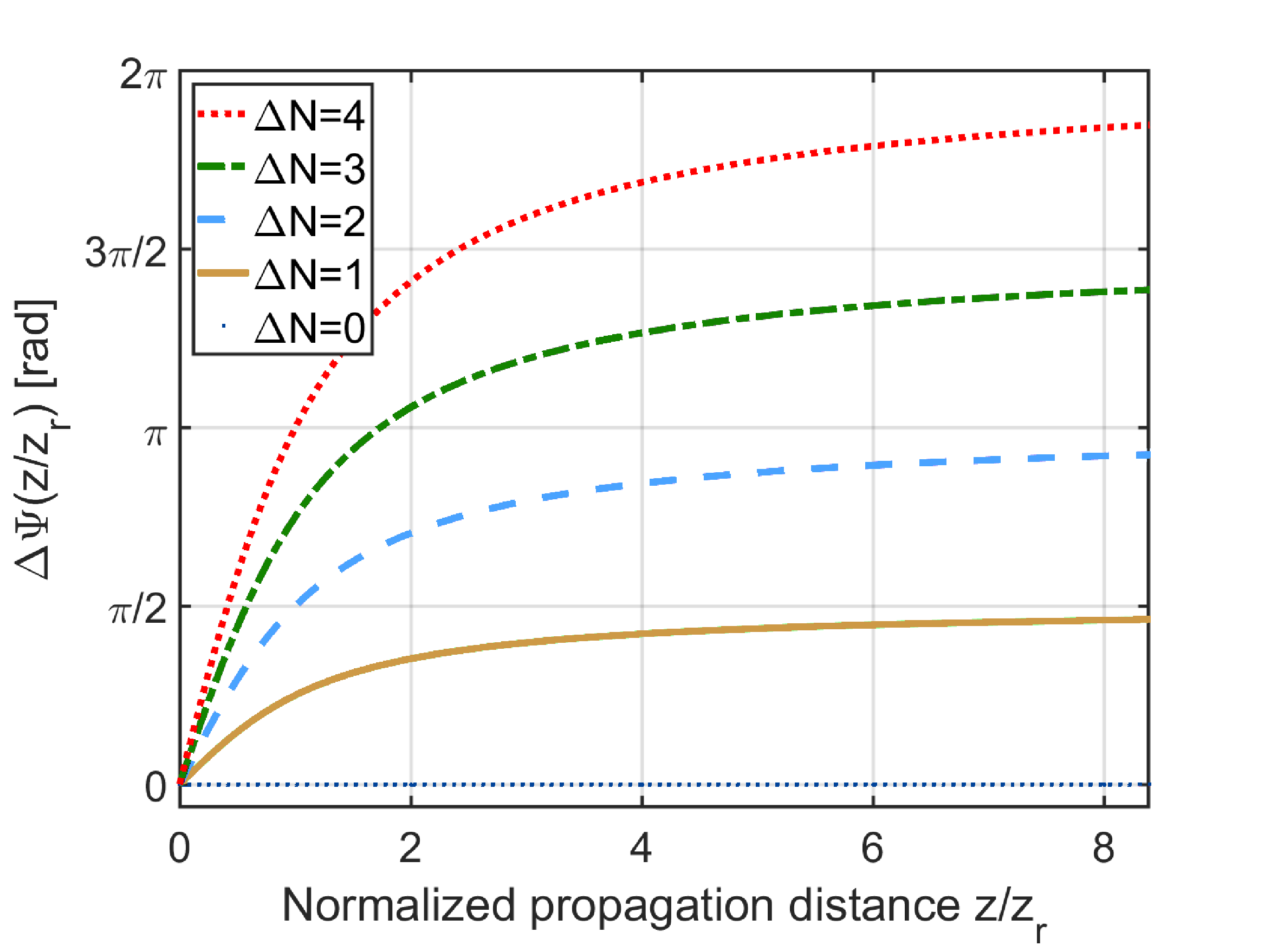}
    \caption{Propagation dependant phase difference $\Delta \Psi(z/z_r)$ of the Hybrid Hermite-Laguerre-Gauss (HHLG) Vector Modes as a function of the normalized propagation distance $z/z_r$ for different mode order differences $\Delta N$.}
    \label{img_deltaModeNumber}
\end{figure}

As stated before, LG and HG modes have phase terms that depend on the propagation distance $z$, $\exp[i\Psi_{L}(z)]$ and $\exp[i\Psi_{H}(z)]$ respectively, where \cite{JO_Rosales_2018}
\begin{equation}
    \Psi_{L}(z)=(2p+|\ell|+1)\tan^{-1}(z/z_r) \equiv (N_{L}+1)\tan^{-1}(z/z_r),
    \label{ec_phaseLG}
\end{equation}
and
\begin{equation}
    \Psi_{H}(z)=(n+m+1)\tan^{-1}(z/z_r) \equiv (N_{H}+1)\tan^{-1}(z/z_r),
    \label{ec_phaseHG}
\end{equation}
Put simply, HHLG vector modes acquire a propagation-dependent phase difference given by
\begin{equation}
    \Delta\Psi(z/z_r) = \Delta N \tan^{-1}(z/z_r),
    \label{ec_intermodalPhase}
\end{equation}
where $\Delta N=N_H-N_L$. For reference, Fig.~\ref{img_deltaModeNumber} shows the phase difference $\Delta\Psi(z/z_r)$ as a function of the normalised propagation distance for different values of $\Delta N$. 

\medskip

HHLG vector modes were experimentally generated through the coherent superposition of two orthogonal scalar modes with orthogonal polarisation states in the circular basis, employing the experimental setup described in \cite{expSetup_2017}. This setup utilises a phase-only Spatial Light Modulator (SLM) and a common path interferometer. Additionally, Stokes polarimetry, as outlined in \cite{Goldstein2011,digitalStokes2020,cox2023}, was employed to measure the polarisation distribution across the transverse plane of the vector modes at various $z$-planes, aiming to characterise their propagation behaviour. To this end, we implemented the digital propagation method described in \cite{digitalProp_2012}, based on the angular spectrum approach \cite{Goodman2017}. This method allows for the calculation of the field distribution of a scalar mode at any arbitrary $z$-plane, $U(\Vec r_\perp,z)$, as:
\begin{equation}
    U(\Vec r_\perp,z)=\mathcal{F}_\perp^{-1}\{\mathcal{F}_\perp\{U(\Vec r_\perp,0)\}\exp(ik_{z}(\Vec{k}_\perp)z)\},
    \label{ec_digitalProp}
\end{equation}
where $\mathcal{F}_\perp\{\cdot\}$ and $\mathcal{F}_\perp^{-1}\{\cdot\}$ denote the two-dimensional Fourier transform and its inverse, respectively; $k_z(\Vec{k}_\perp)$ represents the wave vector component in the propagation direction. Experimentally, the required phase adjustments were applied using the SLM, and the inverse Fourier transform was performed using a biconvex lens with a focal length of $f=250$ mm.


\medskip

\begin{figure}[tb]
    \centering
    \includegraphics[width=0.4\textwidth]{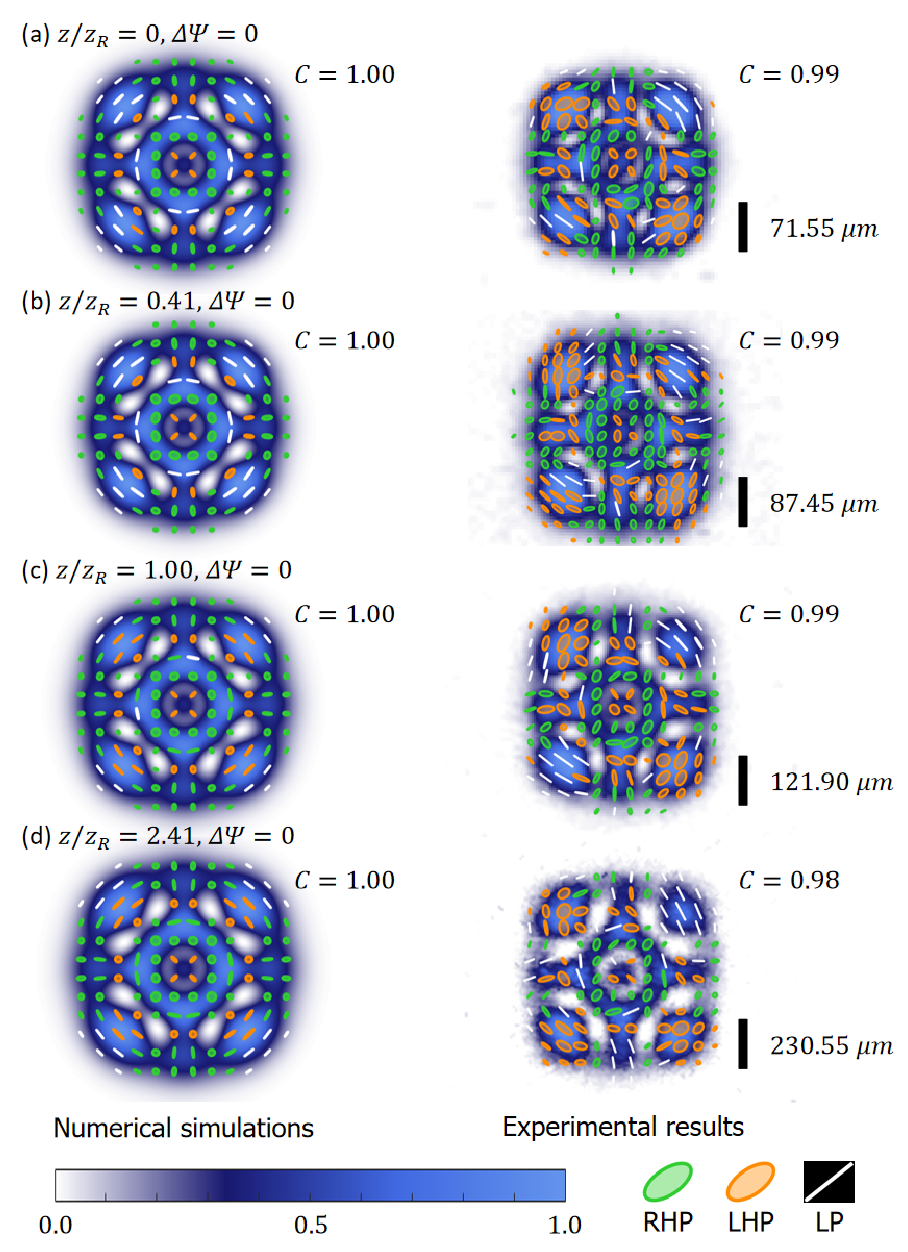}
    \caption{HHLG vector mode for $w_0=0.6$ mm, $\lambda=633$ nm and mode parameters $l=2,\;p=1,\;n=2,\;m=2$ that result in $\Delta N=0$ at different propagation distances: (a) $z/z_r=0$, (b) $z/z_r=0.41$, (c) $z/z_r=1.00$, (d) $z/z_r=2.41$.}
    \label{img_hybridBeamDeltaN=0}
\end{figure}

We generated two examples of HHLG vector modes experimentally and measured their polarisation distributions at various $z$-planes. The first, created through the superposition of $HG_{2,2}$ and $LG^{2}_{1}$ with equal weights ($\theta=\pi/4$ and $\alpha=0$) in Eq.~\ref{ec_vectorHybridBeam}, showed no change in polarisation distribution upon propagation, as evidenced in Fig.~\ref{img_hybridBeamDeltaN=0}. Here, numerical simulations are presented in the left column and experimental results on the right. The beam's size increase due to diffraction is also visible. We computed the concurrence (sometimes called the vector quality factor) \cite{McLaren2015, Selyem2019, Manthalkar2020} to quantify the vector quality of the modes, finding excellent agreement between the simulations and experimental results.

\begin{figure}[tb]
    \centering
    \includegraphics[width=0.4\textwidth]{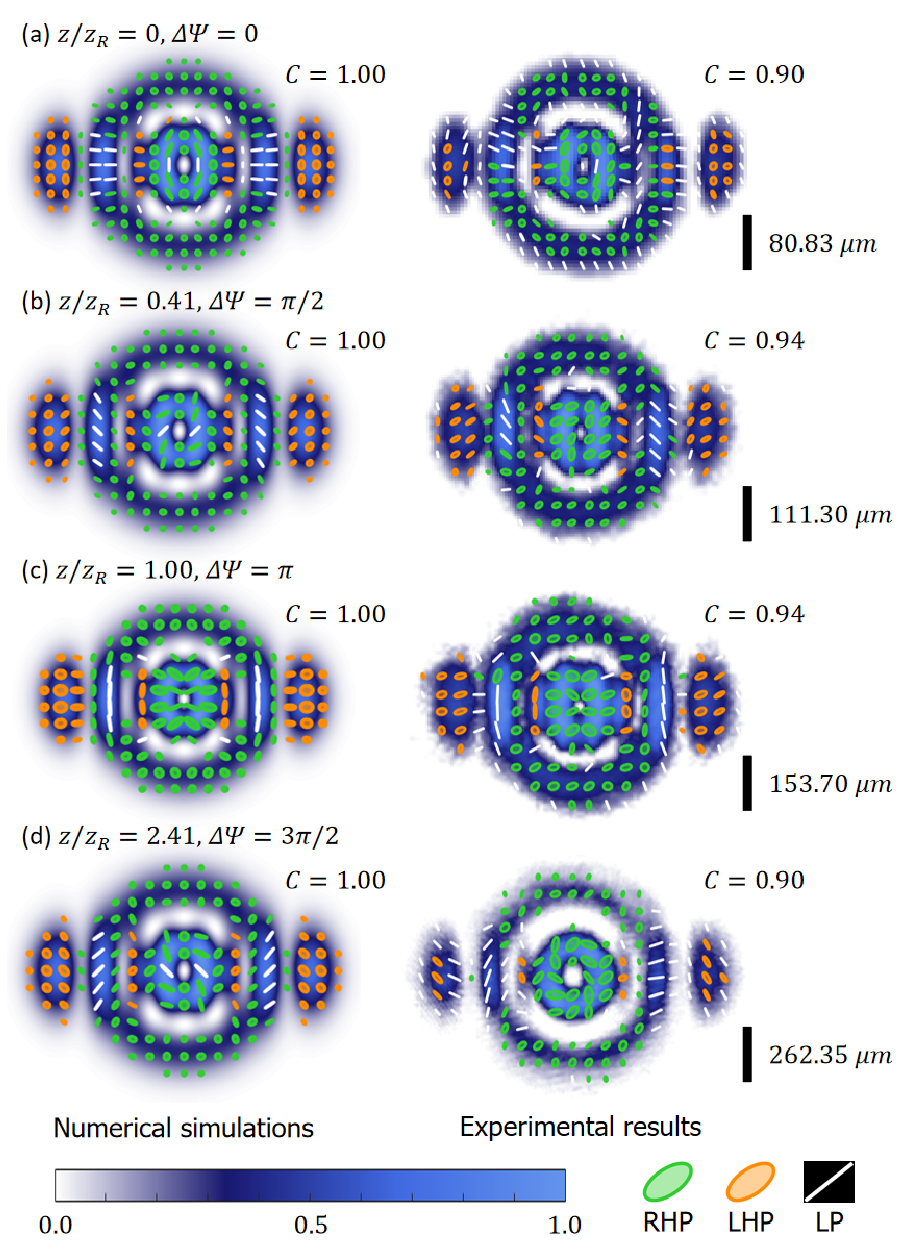}
    \caption{HHLG vector mode for $w_0=0.6$ mm, $\lambda=633$ nm and mode parameters $l=1,\;p=1,\;n=0,\;m=7$ that result in $\Delta N=4$ at different propagation distances: (a) $z/z_r=0$, (b) $z/z_r=0.41$, (c) $z/z_r=1.00$, (d) $z/z_r=2.41$.}
    \label{img_hybridBeamDeltaN=4}
\end{figure}

The second example involved the superposition of $HG_{0,7}$ and $LG^{1}_{1}$ with $\theta=\pi/4$ and $\alpha=0$ in Eq.~\ref{ec_vectorHybridBeam}. The chosen spatial modes resulted in $\Delta N=4$, a scenario not depicted in Fig.~\ref{img_matrixOrthogonal}. We anticipated that this vector mode would acquire a phase difference between the constituent modes upon propagation, as described by Eq.~\ref{ec_intermodalPhase}, corresponding to the red dotted curve in Fig.~\ref{img_deltaModeNumber}. The overlaid polarisation distribution and transverse intensity profile, shown in Fig.~\ref{img_hybridBeamDeltaN=4}, reveal how the linear polarisation states rotate from horizontal at $z=0$ to vertical at $z=z_r$ through propagation.

\medskip

Vector modes of light are becoming increasingly important in contemporary optical laboratories due to their diverse applications and unique fundamental properties. In this study, we aim to contribute to the ongoing research on structured light by introducing a new type of hybrid vector mode. This mode combines the spatial degrees of freedom from two different mode families into a single vector mode. Our calculated superposition has resulted in the Hybrid Hermite-Laguerre-Gaussian (HHLG) vector modes, which show interesting (and configurable) polarisation behaviours during propagation -- perhaps ``knots'' of polarisation? We believe these vector modes could have practical applications in fields such as optical trapping, optical and quantum communications as well as optical metrology, offering new possibilities for exploration and development.

\medskip

\noindent
{\bf Acknowledgements} EMS (CVU: 742790) and LMC (CVU: 894875) acknowledge CONAHCYT for the financial support by means of scholarships for their postgraduate studies. MC would like to acknowledge funding from the South African National Research Foundation. \\

\noindent
{\bf Disclosures} The authors declare that there are no conflicts of interest related to this article.\\

\noindent
{\bf Data availability} Data underlying the results presented in this paper are not publicly available at this time but may be obtained from the authors upon reasonable request.

\end{document}